\documentclass[RRNAAS,twocolumn]{aastex631}
\usepackage{xspace}

\usepackage{cjhebrew}
\usepackage{fontawesome}
\usepackage[T1]{fontenc}
\newfam\hebfam
\font\tmp=rcjhbltx at10pt \textfont\hebfam=\tmp
\font\tmp=rcjhbltx at7pt  \scriptfont\hebfam=\tmp
\font\tmp=rcjhbltx at5pt  \scriptscriptfont\hebfam=\tmp
\edef\declfam{\ifcase\hebfam 
     0\or1\or2\or3\or4\or5\or6\or7\or8\or9\or A\or B\or C\or D\or E\or F\fi}
\mathchardef\tav   = "0\declfam 74

\newcommand{\pz}{photo-$z$\xspace}

\newcommand{\proj}[1]{\textsc{#1}\xspace}
\newcommand{\cosmos}{\proj{COSMOS}}
\newcommand{\lsst}{\proj{LSST}}
\newcommand{\castor}{\proj{CASTOR}}
\newcommand{\euclid}{\proj{Euclid}}
\newcommand{\rst}{\proj{Roman}}
\newcommand{\sfour}{Stage~IV~}

\newcommand{\code}[1]{\texttt{#1}\xspace}
\newcommand{\lephare}{\code{LePhare}}

\defcitealias{Schmidt_2020}{Schmidt \& Malz, et al.}
\newcommand{\citeDCp}{(\citetalias{Schmidt_2020} \citeyear{Schmidt_2020})\xspace}
\newcommand{\citeDCt}{\citetalias{Schmidt_2020} (\citeyear{Schmidt_2020})\xspace}

\begin{document}

\title[TheLastMetric for Stage IV galaxy surveys]{A holistic exploration of the potentially recoverable redshift information of Stage IV galaxy surveys}

\correspondingauthor{Bryan R. Scott}

\author[0000-0002-3894-9823]{Bryan R. Scott}
\affiliation{Center for Interdisciplinary Exploration and Research in Astrophysics (CIERA), 
Northwestern University, 
1800 Sherman Ave, 8th Floor, Evanston, IL 60201}
\affiliation{LSST Discovery Alliance Data Science Fellowship Program}
\email{bryan.scott@northwestern.edu}

\author[0000-0002-8676-1622]{Alex I. Malz}
\affiliation{McWilliams Center for Cosmology and Astrophysics, Department of Physics, Carnegie Mellon University}

\author{Robert Sorba}
\affiliation{Institute for Computational Astrophysics, Saint Mary’s University, 923 Robie Street, Halifax, Nova Scotia B3H 3C3, Canada}

\begin{abstract}

Extragalactic science and cosmology with Stage IV galaxy surveys will rely almost exclusively on redshift measurements derived solely from photometry, which are subject to systematic and statistical uncertainties with numerous analysis choices, such as that of an estimator and prior information and no universal solution. Single-survey photometric redshift estimates ought to be improved by combining data from multiple surveys, with common wisdom asserting that optical data benefits from additional infrared coverage but not from additional ultraviolet coverage. The degree of improvement for either case is not well-characterized and attempts to quantify it necessitate assumptions of a chosen estimator and its prior information. We apply an information theoretic metric of potentially recoverable redshift information to assess the impact of multi-survey photometry combinations without assuming an estimator nor priors in the context of the Vera C. Rubin Observatory Legacy Survey of Space and Time (LSST) in the optical, Roman and Euclid Surveys in the infrared, and Cosmological Advanced Survey Telescope for Optical-UV Research (CASTOR) in the ultraviolet. We conclude that the addition of UV photometry can benefit redshift determination of certain galaxy populations, but that gain is tempered by their decreased chance of meeting detection criteria at higher wavelengths. We explore the spectral energy distribution (SEDs) of galaxies whose potentially recoverable redshift information is most impacted by additional photometry to provide a basis for future evaluation of which science cases may motivate which combinations. The holistic assessment approach we develop here is generic and may be applied to quantify the impact of combining photometric datasets, changing experimental design choices, including observing strategy optimization, and evaluating systematics mitigation procedures.

\end{abstract}


\section{Introduction}
\label{sec:intro}

Stage IV galaxy imaging surveys, including the Vera C. Rubin Observatory Legacy Survey of Space and Time (\lsst), Roman Wide Field Survey, and Euclid Survey, will produce large catalogs of multi-band images, corresponding photometry, and source positions. 
As a consequence of the scale and depth of these surveys, distance information in the form of spectroscopic redshifts will be unavailable for all but a small fraction of the catalogs. 
Instead, these surveys will rely on photometric redshift (photo-$z$) estimates derived from the mapping of features of galaxy spectral energy distributions (SEDs) through each survey's filter set (\cite{Hildebrandt_2021}, \cite{Crenshaw_2020}, \cite{Brammer_2008}). 
Stringent accuracy and precision requirements on the characterization of this mapping has led to substantial effort across the community to improve the quality of photo-$z$ estimates, and it remains one of the most pressing priorities for ensuring the success of these experiments across extragalactic science and cosmology. 

Even in light of these ongoing efforts, uncertainties in photo-$z$ estimates due to prior information imperfection or physical degeneracies in the relationship between galaxy colors and redshifts are likely to contribute to the systematics floor on extragalactic science from these surveys. 
Efforts to characterize this relationship include improvements to estimators (cf. \cite{BNN_photo_z}, \cite{RF_photo_z}, \cite{BPZ}) and improvements to training sets (cf. \cite{augment_training}, \cite{Masters_photo_z}).  
\cite{Graham_2020} have identified the addition of near-ultraviolet (UV) photometry and near-infrared (IR) photometry as an avenue to improving aggregate photo-$z$ performance by breaking degeneracies present in optical surveys alone, highlighting degeneracies between colors derived from the shifting of the Lyman and Balmer breaks through ground based optical photometry at $z \approx 0.2$ and $z \approx 1.5$ in the \lsst filter set.
Despite these results and the importance of similar degeneracy breaking in the near-infrared, the shifting of the 4000 $\AA$ break into the IR at high redshifts has been argued to limit the value of optical-UV photometry for photo-z estimation.

This paper expands upon past work (for reviews, see \cite{Newman_2022} and \cite{Brescia2021}) in assessing photo-$z$ quality from single and combinations of future Stage IV surveys, specifically assessing the impact of IR versus UV photometry. 
We consider the Cosmological Advanced Survey Telescope for Optical-UV Research (\castor) Wide-Field Survey, which is the deepest, widest, and highest resolution wide field UV imaging survey proposed for the late 2020s and early 2030s \citep{2024AJ....167..178C, 2012SPIE.8442E..15C}.
As presently envisioned, CASTOR will produce a deep survey ($\ge$ AB mag 27) in the $u$, $uv$, and $g$ over $\approx 5-10 \%$ of the sky over the lifetime of the mission to complement the Stage IV cosmology surveys we consider: 
the Rubin \lsst Wide Fast Deep Survey \citep{lsstsciencecollaboration2009lsst}, \euclid \citep{laureijs2011euclid}, and the \rst High Latitude Survey \citep{EiflerHLS}. 
A secondary goal of this paper is to understand how photo-$z$ recovery quality varies for specific galaxy SED populations. 
These goals are motivated by recent improvements in photometric redshift metrics that focus on the information content of photometry rather than the performance of specific estimators under pre-survey conditions.

Historically, photo-$z$ recovery performance was quantified through summary metrics (cf. \cite{Graham_2017}) used to inform survey design and analysis choices, such as whether to focus effort on improving estimators and optimally allocating spectroscopic observations. 
Unfortunately, it is difficult to make these decisions based on traditional metrics, such as the mean bias or outlier fraction, because they are dependent on assumptions about the technique used to characterize the color-redshift relation, the specific science case being studied, and are moreover point estimates of the redshift distribution that can not be fully propagated into astrophysical or cosmology inference. 
Even in the artificial case where metrics are measured for multiple algorithms on simulated datasets using a shared parameterization of the photo-$z$ distribution $n(z)$, interpretation remains challenging since implicit prior information is not, by definition, shared across estimators \citeDCp.

To address these limitations, \cite{malz2021informationbased} introduces \textit{TheLastMetric}, an information theory-based approach that approximates a lower bound on the mutual information between redshift and photometry in the joint probability distribution thereof, implemented using \code{pzflow} \citep{crenshaw2024probabilistic}. 
The application of \textit{TheLastMetric} to observing strategy optimization builds on earlier applications of information theory to derived data product storage in \citet{Malz_2018} and photometric filter design in \cite{Kalmbach_2020}.
Because this technique pertains to the joint distribution of redshifts and photometry, it yields metrics of photo-z performance that are i) in shared units between surveys and downstream science use cases, and that ii) do not depend on a choice of photo-$z$ estimation method nor its associated assumptions, e.g.~ knowledge of the redshift distribution $n(z)$. 
Instead of quantifying the quality of photo-$z$ estimates from some estimator and assumptions, \textit{TheLastMetric} may be understood as a measure of the potentially recoverable information about redshift contained in some photometric data set.

This paper is organized as follows. 
We begin in Section~\ref{sec:data} by describing our process for generating a conditional density distribution of redshift on photometry derived from empirical catalogs. 
In Section~\ref{sec:meth}, we introduce the information theoretic metrics we apply. 
In Section~\ref{sec:res}, we present the result of applying our approach to future surveys individually and in combinations and investigate some surprising observations by way of populations of galaxy SEDs. 
In Section~\ref{sec:conclusions}, we conclude with a prospectus on the role of combined broadband photometry for extragalactic astronomy and cosmology in the coming decade.

\begin{figure}
  \centering
  \includegraphics[width=0.5\textwidth]{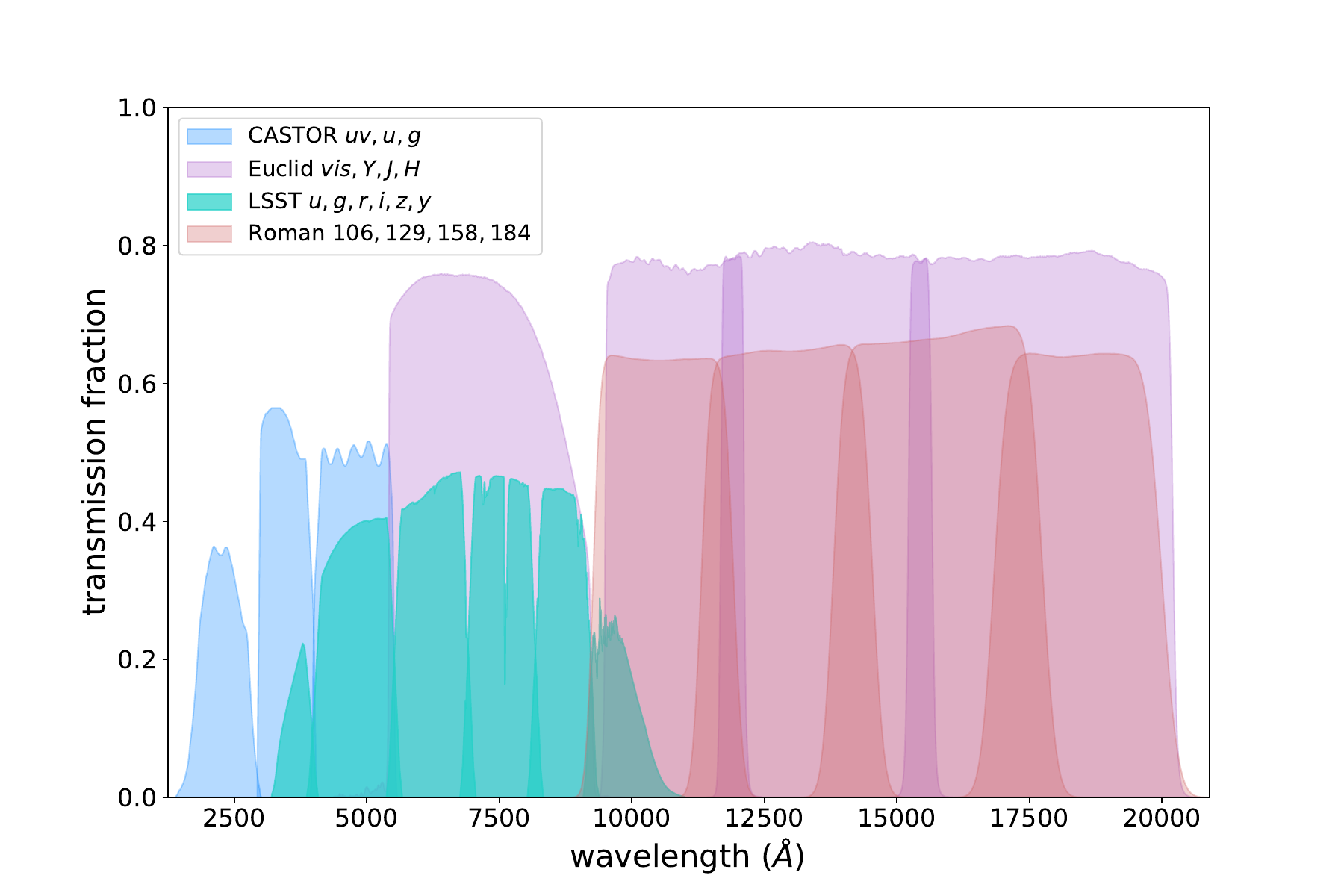}
  \caption{Photometric filter throughput curves of the four surveys considered.
  }
  \label{fig:filts} 
\end{figure}

\section{Data}
\label{sec:data}

Our formalism for measuring the redshift information content of a photometric survey requires simulated redshifts and photometry. 
Section~\ref{sec:sims} describes the production of mock catalogs based on \cosmos2020 galaxy SEDs.
Section~\ref{sec:cats} describes the definition of samples used in this study reflecting the selection functions of \euclid, \lsst, \rst, and \castor.

\subsection{Mock galaxy catalog simulation}
\label{sec:sims}

We begin by generating mock photometric observations using the \cosmos2020 Farmer catalog, release 1, version 2.0 \citep{Weaver_2022}. 
From the initial sample of 964,506 sources, only 698,211 galaxies are well-fit (unflagged) by the \lephare \citep{LePhare_code} spectral templates and have estimates for the median stellar mass and star formation rate (SFR) are used (\texttt{FLAG\_COMBINED == 0, lp\_type == 0, lp\_model > 0, lp\_NbFilt > 5, !(lp\_mass\_med.mask || lp\_SFR\_med.mask}). 
We treat the Farmer catalog photometry and best-fit \lephare parameters (redshift, stellar mass, and SFR) as the ``true'' underlying properties of each galaxy from which to build mock photometric catalogs for the Stage IV surveys.

To determine the mock photometric measurements for \euclid, \lsst, \rst, and \castor, a color correction method is used. 
For each galaxy and for each bandpass shown in Figure~\ref{fig:filts}, color corrections from similar wavelength \cosmos bandpasses are found using the best-fitting \lephare spectral template (corrected for dust and IGM absorption). 
The photometric magnitude in a mock bandpass for a galaxy is then the weighted average of the \cosmos measurements plus the color correction term. 
Using this approach places less significance on the exact choice of spectral template model, since the correction term from nearby bandpasses (in wavelength) should be small regardless of which templates are used. 
Additionally, any magnitudes that are not well reproduced by the \lephare templates are preserved so that our mock magnitudes more closely resemble those from the \cosmos catalog, rather than perfectly matching the (ill-fitting) template. 
When there are no observations at similar wavelengths to the mock bandpass, the mock magnitude is found in the typical way of integrating the scaled template model flux through the bandpass transmission curve. 
This process yields mock photometry for future surveys which are again treated as a ``reality'' from which to simulate noisy observations.

Accurate noise estimates require knowledge of the angular size of each galaxy. 
For each galaxy, we estimate the half-light radius in kpc using the average mass-to-size relation and intrinsic dispersion from \citet{Kawinwanichakij_2021}. 
The physical size is converted to an angular full width at half maximum (FWHM) using a flat $\Lambda$CDM cosmology with $H_0 = 0.7$ and $\Omega_{m0} = 0.3$. 
The galaxy's FWHM is convolved with each telescope's point spread function (PSF) and an observation aperture diameter of 1.346 times the convolved FWHM (see Eq. 19 of \cite{LSE40} (\lsst LSE-40)), resulting in the determination of the number of pixels ($N_{pix}$) in the aperture. 
The signal-to-noise ratio depends on the instrument parameters and is given by
\begin{equation}
\label{eq:snr}
SNR = \frac{e_{source}}{  \sqrt{e_{source} + e_{sky} + e_{dark} + N_{pix}N_{exp}\sigma^2_{read}} } ,
\end{equation}
where $e_{source}$ and $e_{sky}$ are the total electron counts in the aperture from the galaxy source and the sky respectively, found using the total exposure time and the magnitude zero points, $e_{dark}$ is the total number of electrons generated from the dark current parameter multiplied by $N_{pix}$ and the total exposure time, and $\sigma_{read}$ is the read noise in electrons per exposure.

Mock observations of each galaxy for the Stage IV surveys are finally obtained by randomly perturbing the assumed true underlying photometry in flux-space using the noise as the standard deviation of a normal distribution.
The final result is a pseudo-empirical catalog of mock photometric observations for approximately 700,000 galaxies drawn from the \cosmos2020 Farmer catalog. 
The mock observations thus mimic reality as closely as possible, and are limited only by the depth and wavelength coverage of the \cosmos survey.

\begin{table}
    \hspace{-0.5cm}
    \begin{tabular}{|l|}
    \hline
    \castor\\
    \hline
    $uv < 27.5$ \\
    $u < $ 27.5\\
    $g < $ 27.6\\
    \hline
    \end{tabular}
    \hspace{-0.5cm}
    \begin{tabular}{|l|}
    \hline
    \lsst\\
    \hline
    
    $u < 25.6$ \\
    $g < 26.9$ \\
    $r < 26.9$ \\
    $i < 26.4$ \\
    $z < 25.6$ \\
    $Y < 24.8$ \\
    \hline
    \end{tabular}
    \hspace{-0.5cm}
    \begin{tabular}{|l|}
    \hline
    \euclid\\
    \hline
    
    $v < 26.2$ \\
    $y < 24.5$ \\
    $j < 24.5$ \\
    $h < 24.5$ \\
    \hline
    \end{tabular}
    \hspace{-0.5cm}
    \begin{tabular}{|l|}
    \hline
    \rst\\
    \hline
   $106 < 26.7$ \\
   $129 < 26.7$ \\
   $158 < 26.7$ \\
   $184 < 26.7$ \\
    \hline
    \end{tabular}
    \caption{The limiting magnitudes of each photometric filter considered for the mock survey sample definitions.}
    \label{tab:limits}
\end{table}

\begin{figure}
  \centering
  \includegraphics[width=0.5\textwidth]{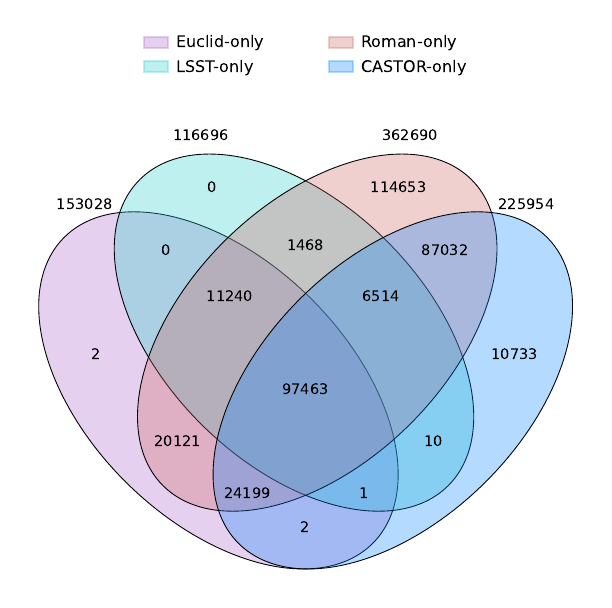}
  \caption{The number of mock galaxies in each simulated survey and all combinations.
  }
  \label{fig:venn} 
\end{figure}

\subsection{Sample definition}
\label{sec:cats}

As our mock galaxies are based on \cosmos photometry much deeper than any of the wide-field \sfour surveys, we first define a subsample based on the expected per-band limiting magnitudes, provided in Table~\ref{tab:limits}, for each of the surveys and survey-combinations we consider, with resulting catalog sizes shown in Figure~\ref{fig:venn}. 
We impose the conservative requirement that included galaxies must meet the detection threshold in every band. 
While this selection criterion significantly reduces the total number of detected sources in a combined catalog, it saves us from having to assume a forced photometry procedure for non-detections.

While the analysis of all combinations of the four surveys could be performed, for simplicity of the presentation of the results, we consider only a subset of combinations inspired by Figure~\ref{fig:filts}.
Due to the significant overlap of the \euclid and \rst filters, it is sufficient to use one, arbitrarily chosen to be \euclid, except when the two are considered in combination with each other.

\begin{figure*}
  \centering
  \includegraphics[width=\textwidth]{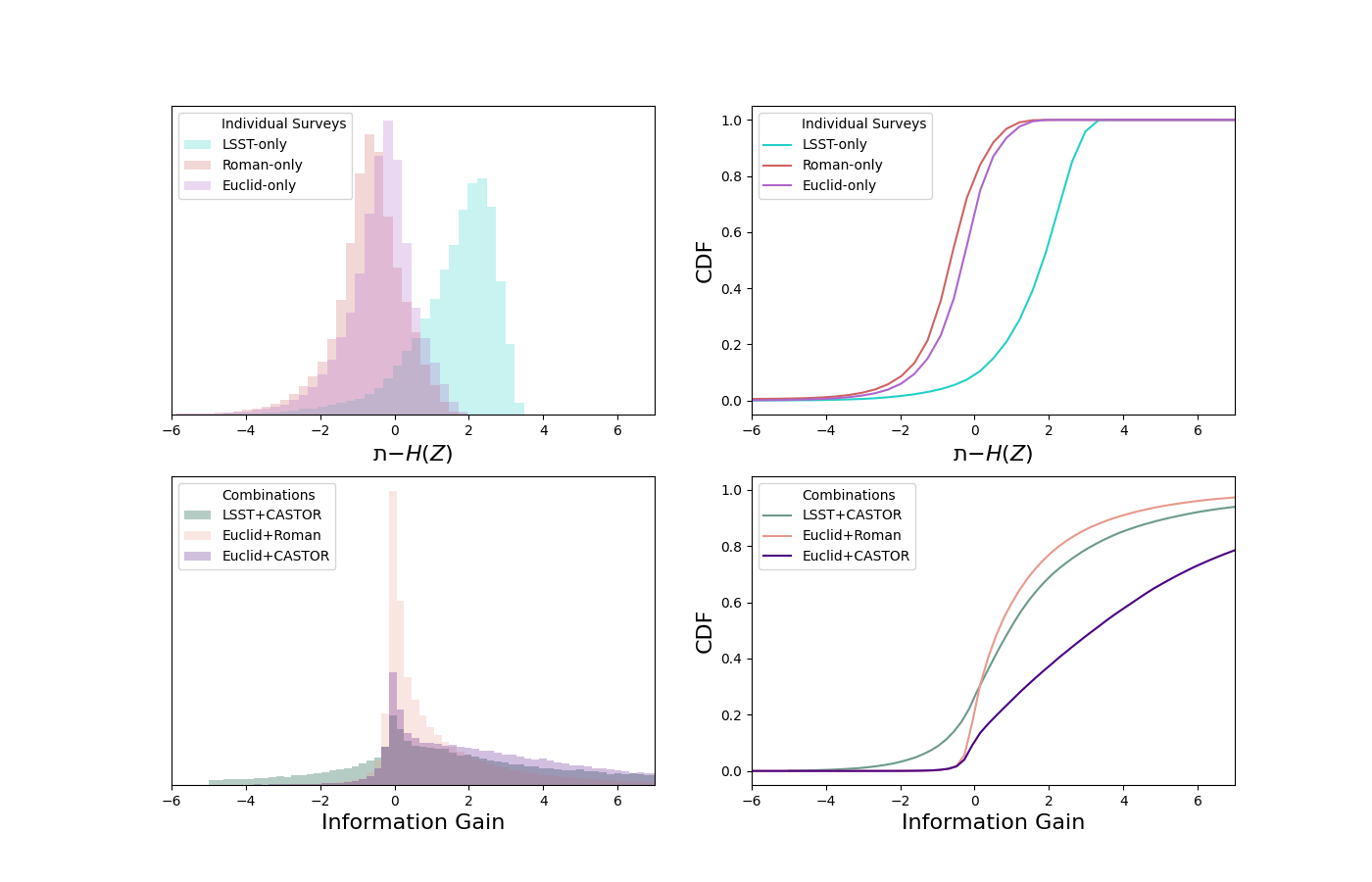}\\
  \caption{Top left: distributions of the redshift information content of \lsst, \rst, and \euclid.
  Top right: cumulative distribution function of top left.
  Bottom left: distribution of point-wise information gain relative to single-survey observations.
  Bottom right: cumulative distribution function of bottom left.
  On a population level, \lsst photometry is already more informative than \rst or \euclid photometry alone. 
  The point-wise information gain is really the difference in information relative to the single-survey case, so it may be negative.
  \lsst+\castor has a much more symmetric distribution than the other combinations, meaning the information content is diluted by the additional UV photometry about as often as it is enhanced by it.
  The strong peaks at zero and long tails of the \euclid+\rst and \euclid+\castor combinations show that while many galaxies don't experience a significant gain in information due to the additional photometry, a significant fraction also do see a major gain, and very few galaxies have their information content diminished by the additional photometry.
 }
  \label{fig:tavdists} 
\end{figure*}

\section{Methods}
\label{sec:meth}

In order to characterize the photo-$z$ quality of Stage IV surveys individually and in combinations, we require metrics that are independent of the choice of estimator and associated priors and that can be compared across surveys. 
Section~\ref{sec:info} employs information theory to motivate the base metric around which we build our analysis, and Section~\ref{sec:gains} presents the summary statistics of this metric we explore in this study.  

\subsection{Redshift mutual information metric}
\label{sec:info}

Traditional \pz performance metrics are demonstrated throughout the literature and presented comprehensively in \citeDCt. 
The ingredients for such metrics are photometric data with known redshifts (either real data with spectroscopic confirmation or mock data with true redshifts), an implementation of each estimator, and prior information for the estimator(s) (either training sets for empirical methods or template libraries for model-based methods).
Making appropriate choices for the last two of these in particular is challenging due to the fast pace of methodological development in this field and complexity of coordinating sharing of spectroscopic follow-up data.
To avoid baking such assumptions into future analysis design choices, one can consider information theoretic metrics such as \citet{malz2021informationbased, Kalmbach_2020, Malz_2018}, which have been applied to the optimization of observing strategy, photometric filter design for Balmer peak localization, and derived data product storage format, respectively.
While both classes of metrics assume some notion of an underlying space of redshifts and photometry, which of course cannot be known before the survey is conducted, avoiding the assumptions of an estimator and prior information provides some protection from the risk of optimizing analysis decisions for the methods and priors of an earlier era. 

We employ TheLastMetric\footnote{$\tav$, pronounced "tav", is the last letter of the Hebrew alphabet.} \citep{malz2021informationbased}, defined as
\begin{equation}
\label{eq:tav}
    \tav \equiv E_{p(z, x_{phot})} \log q_{\phi} (z, x_{phot}) + H(z) \leq I(z; x_{phot}) ,
\end{equation}
a variational approximation to the lower bound of mutual information $I(z; x_{phot})$ between redshift $z$ and photometry $x_{phot}$, which quantifies the decrease in uncertainty on redshift due to knowing the photometry, where $H(z)$ is the entropy or self information, computed directly from the marginal redshift distribution $p(z)$ in the simulated catalogs, $q_{\phi}$ is the approximation to the joint probability distribution $p(z, x_{phot})$ and $E$ represents the expectation value.

We use the \code{TheLastMetric} implementation of a normalizing flow fit to a mock catalog of redshift and photometry, which is built on \texttt{pzflow} \citep{Crenshaw_2024}.

In order to account for epistemic or model uncertainties, we use an ensemble of 30 randomly initialized normalizing flows, each trained for 150 epochs, in which the data undergoes a permutation at each step, generously exceeding the empirically determined rate of training loss convergence.

\subsection{Summary statistics of information gain}
\label{sec:gains}

Distributions of $\tav$ in redshift- or color-space and differences thereof between data sets can be used to characterize the redshift information content of photometry from different survey combinations. 
To test our hypothesis that degeneracy-breaking induces an increase in the redshift information content of photometry in a combined survey, we construct an expression compare the mutual information $I(z; x_{phot})$ between an individual survey and combination thereof.

The mutual information can be broken down as
\begin{equation}
\label{eqn:mutual}
    I(z; x_{phot}) = H(x_{phot}) - H(x_{phot}|z) 
\end{equation}
in terms of the entropies $H(z)$ and $H(z|x_{phot})$, where the conditional entropy is
\begin{equation}
\label{eqn:entropy}
    H(x_{phot}|z) = - E_{p(z, x_{phot})} \log[p(z|x_{phot})]. 
\end{equation} 
Taking the difference of Equation~\ref{eqn:mutual} for the data sets corresponding to a single-survey $x_{phot,S}$ and a combination of surveys $x_{phot,SC}$ gives

\begin{eqnarray}
\Delta I(z; x_{phot, SC}) =  
E_{q(z, x_{phot,SC})} \log[q_{\varphi}(z|x_{phot,SC})
\nonumber \\ - E_{p(z, x_{phot,S})} \log[p_{\varphi} (z|x_{phot,SC}], 
    \label{eqn:deltaI}
\end{eqnarray}
which can be interpreted as the gain in mutual information due to using a combination of surveys rather than just one.

Note that Equation~\ref{eqn:deltaI} does not depend on the entropy of the $x_{phot}$, nor is it positive definite, allowing the mutual information gain to be negative if additional photometry adds more noise than redshift information to estimates of the conditional distribution. 

Figure~\ref{fig:tavdists} illustrates a cursory comparison of single-survey $\tav$ distributions and the differences defined in Equation~\ref{eqn:deltaI} for a few survey combinations, providing both a histogram of the metric and its cumulative distribution function (CDF) for each data set. 

As anticipated, \euclid and \castor both have distributions that peak at slightly negative values of $\tav - H(z)$ (that is, at the expectation of the log-joint posterior of redshift on photometry). 
\lsst photometry is typically much more informative, with a distribution that peaks at a much larger, and positive value, reflecting our expectation that the redshift information in optical photometry is conveyed primarily by the locations of the spectral breaks as they pass through the photometric filters.

The distribution of per-galaxy information gain for \euclid+\rst (compared to \euclid-only) is strongly peaked around zero, with non-zero point-wise information gain values lying between 0 and $\approx 3$. 
The addition of \castor to \euclid only sees positive point-wise information gain, but with much longer tails towards high values, a beneficial consequence of the wide separation between bandpasses for this combination.

On the other hand, the addition of \castor to \lsst leads to a large proportion of negative values of $\Delta I(z,x_{ph})$;
because \lsst-detected sources already have some coverage in \castor's wavelength range, the inclusion of shallower measurements thereof adds more noise than new redshift information.

\begin{figure*}
  \includegraphics[trim=0cm 0 0 -1cm, scale=0.5]{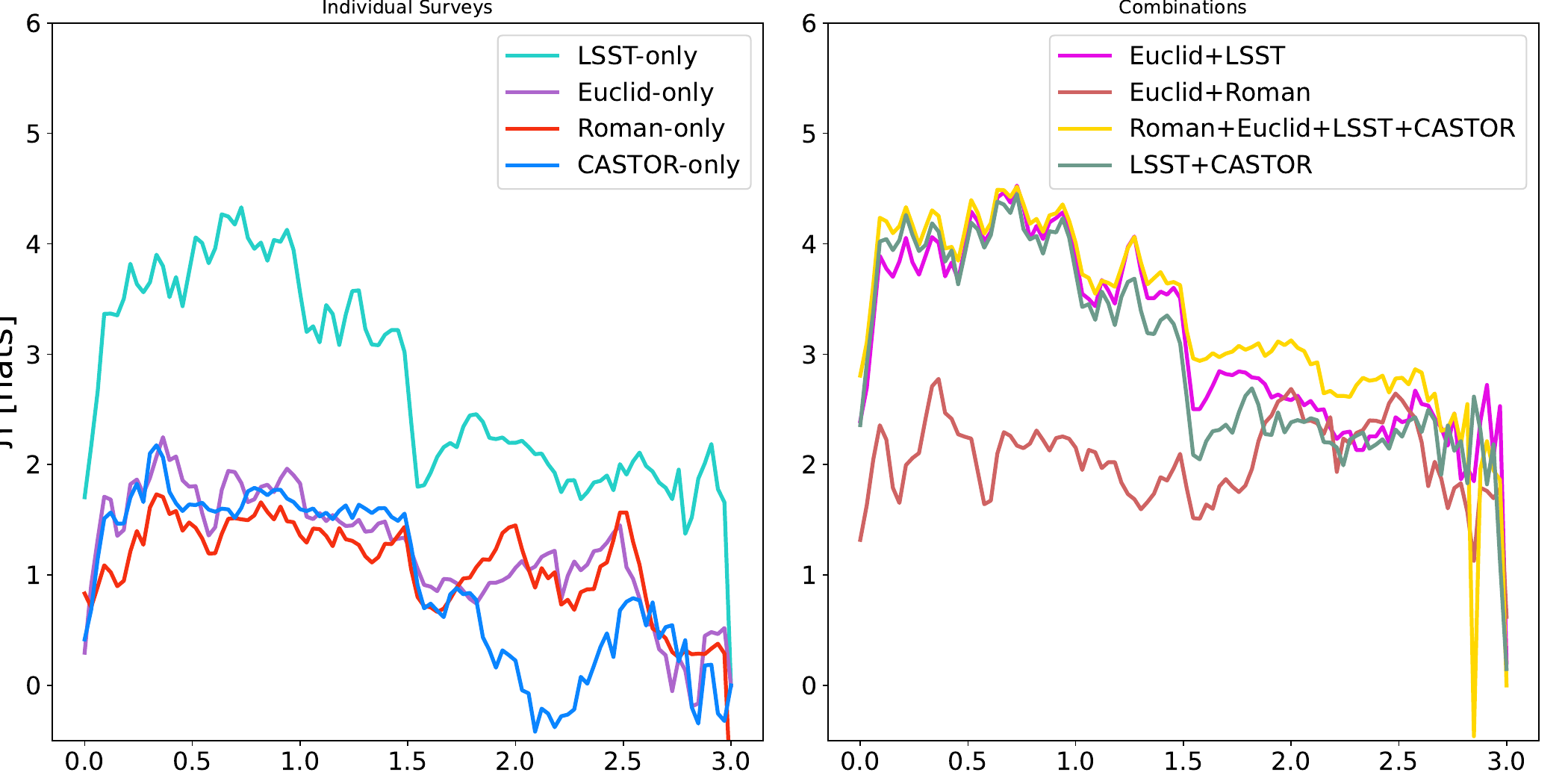}
  \caption{
  Redshift-binned average of $\tav$ for the Stage IV photometric surveys (left) and select combinations thereof (right) in bins of true redshift. At all redshifts, \lsst has more potentially recoverable redshift information than the other surveys, but there are some areas of redshift-space where other surveys contribute distinct information.
  } 

  \label{fig:tlmz} 
\end{figure*}

\begin{figure}
  \includegraphics[scale=0.44]{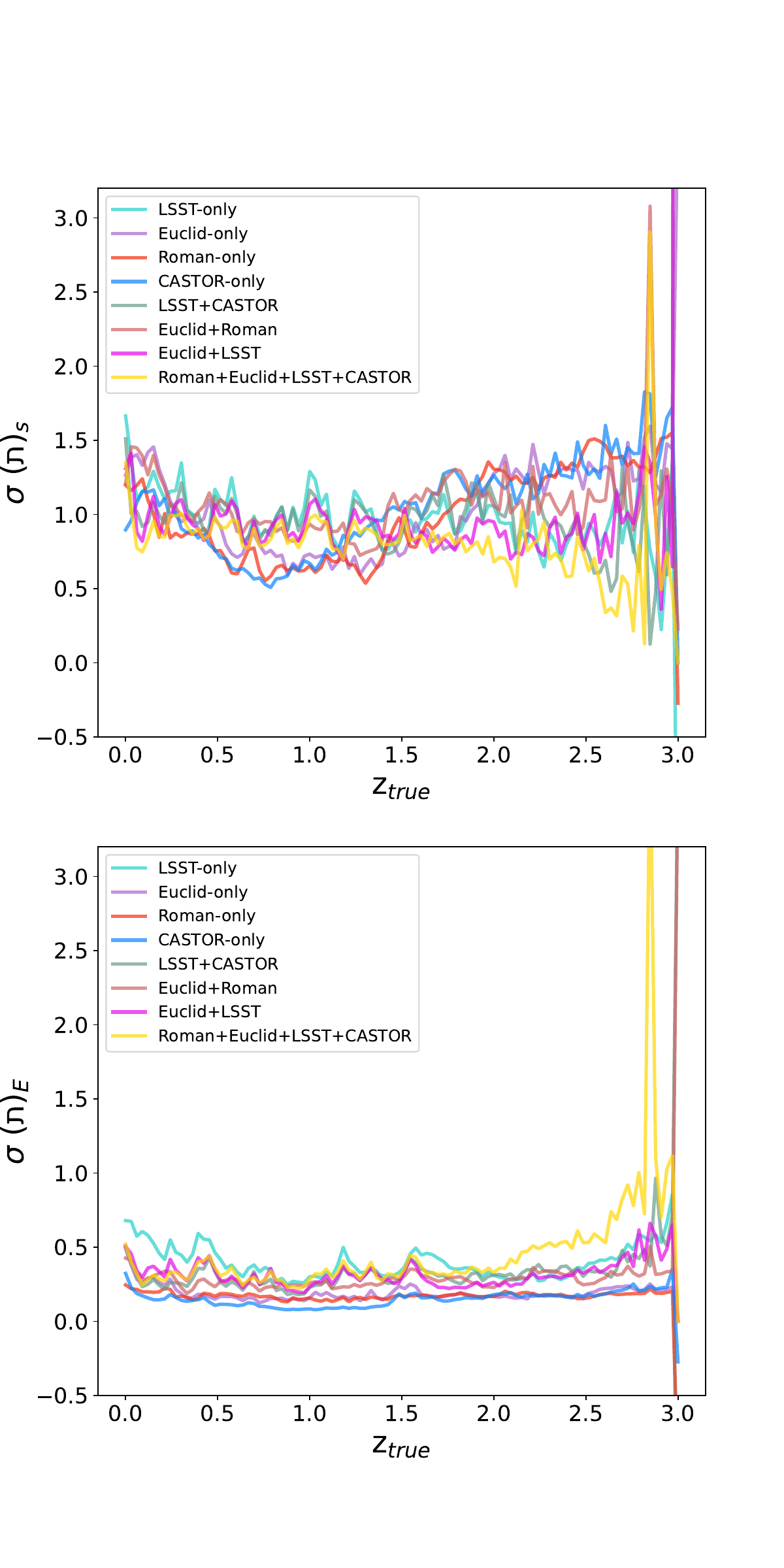}
  \caption{
  Top panel: Statistical uncertainty given as the standard deviation of the values of $\tav$ computed on a per galaxy basis for each catalog. Bottom panel: Epistemic uncertainty, as the standard deviation of the mean values in $\tav$ computed over the ensemble of normalizing flows. All survey combinations we consider show a small peak in the epistemic uncertainty at around z $\approx$ 1.6 as expected from the Lyman-Balmer degneracy, an effect which is larger prior to incorporating the survey selection functions.
  }
  \label{fig:tlmE} 
\end{figure}

\section{Analysis \& discussion}
\label{sec:res}

Here we present the in-depth analysis of the redshift information content of Stage IV surveys.
Section~\ref{sec:zdep} presents the evolution of $\tav$ with redshift for the surveys individually and in combinations. 
Section~\ref{sec:uvinfo} tests our hypotheses regarding the degeneracy-breaking potential of near optical and UV photometry. 
In Section~\ref{sec:tavcolor}, we explore the color space distribution of $\tav$.
Section~\ref{sec:tavselect} applies redshift information gain as a discriminator of galaxy populations to better understand some initially unexpected results. 

\subsection{Redshift dependence of \textit{TheLastMetric}}
\label{sec:zdep}

The redshift dependence of the information theoretic metric $\tav$ for a given survey or survey combination depends on the locations of broadband SED features and spectral breaks relative to the photometric filters. 
In particular, analyses combining multiple surveys benefit from the shift of an SED feature from the bandpass of one survey into another, for example, from the optical \lsst bands into infrared \euclid or \rst filters. 
For a single survey, we therefore expect to see $\tav(z)$ rise at an intermediate redshift where SED features are localized in a given survey's bands (and therefore color gradients are largest) and drop where two SED features are degenerate with one another. 
At redshifts where the galaxy population has informative photometry in a single survey, the information may saturate and not benefit from additional measurements in the filters of another survey.

Figure~\ref{fig:tlmz} shows redshift binned estimates of the mutual information lower bound $\tav$ for individual surveys and combinations thereof. 

Among the single survey cases, \lsst photometry contains the most information at all redshifts, peaking around $z \approx 0.6$. 
At lower redshifts $z \le 1.5$, \euclid, \rst, and \castor have comparable redshift information that drops at higher redshift, moreso for \castor than for the IR surveys. 

All four surveys show a significant dip at the Lyman-Balmer break degeneracy redshift of $z \approx 1.6$. 

We see evidence for the expectations reported in \cite{Graham_2020} that \euclid and \rst should benefit from detecting the Balmer break in Y-J at $z \approx 1.5$ and H-K at $z \approx 2.5$ where both have peaks in $\tav(z)$ before declining at higher redshift.

Next, we examine the different combinations of surveys. 
Given how similar $\tav(z)$ looks for \euclid and \rst individually, it is surprising to see that their combination differs so much from each alone;
in other words, neither survey completely saturates the redshift information available in the IR, which allows the combination to have a larger redshift information content throughout the IR.
Above the Lyman-Balmer degeneracy ($z \gtrapprox 1.5$), the inclusion of IR photometry increases the redshift information relative to \lsst alone, as expected. 
By contrast, \lsst saturates the information content at $z \le 1-1.5$, showing essentially no improvement with the addition of UV and IR photometry,

contradicting the expectation reported in \cite{Kalmbach_2020} that using information theory to optimize filters for \pz s corresponds to designing filters that can optimally constrain the location of the Balmer break, a matter we will discuss more deeply in Section \ref{sec:uvinfo}.

We isolate the statistical and epistemic uncertainty that comprise the reported standard deviation in Figure~\ref{fig:tlmE}, showing the average in redshift bins across the galaxy population and within the flow ensemble, respectively. 
The former serves as an estimate of the statistical uncertainty in the information metric which arises due to sampling of the galaxy distribution, while the latter captures the model uncertainty arising from using independent realizations of the normalizing flow. 

Statistical uncertainties are about a factor of two higher than epistemic/model uncertainty and exhibit roughly flat evolution with redshift owing to the fact that the width of the scatter in galaxy color doesn't change with redshift. 
Although there is a peak in the epistemic uncertainty around the Lyman-Balmer degeneracy for combinations with \lsst, the model uncertainty shows only a minor feature corresponding to this degeneracy at $z_{true} \approx 1.5$, as well as several hints of similar feature-based filter-dependent degeneracies at lower redshifts, which may correspond to high-scatter populations found using traditional metrics in the $z_{phot}-z_{true}$ plane that arise due to uncertainties in the location of the Balmer break.

\subsection{Information content of UV photometry}
\label{sec:uvinfo}

\begin{figure}
  \includegraphics[width=0.5\textwidth]{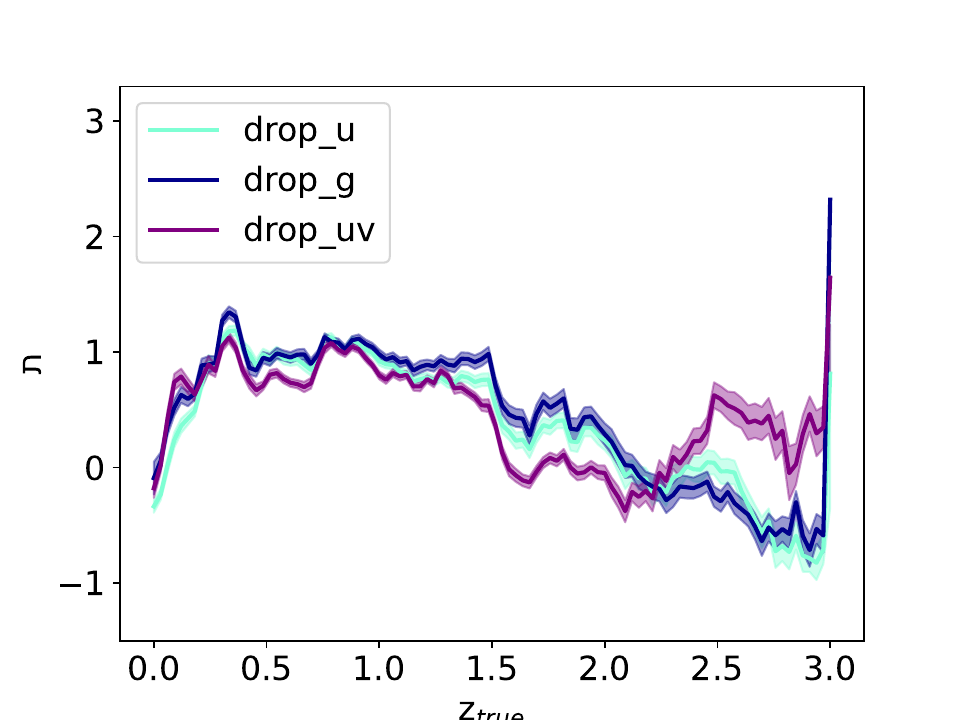}
  \caption{Redshift evolution of $\hat{\tav}$, with standard deviation over the input catalog, for a fixed photometric catalog but dropping each of the three \castor filters. A smaller value of $\hat{\tav}(z)$ corresponds to a filter that contributes more to the conditional information of photometry on redshift. 
  At $z \gtrsim 2$ the uv becomes less informative than the u and the g bands. }
  \label{fig:drop} 
\end{figure}

Because \citet{Graham_2020} predicted that increased \lsst u-band depth would improve the bias and scatter of photo-$z$ estimates more than adding \euclid photometry (but that the addition of \euclid photometry would reduce the $z \ge 1$ catastrophic outlier fraction compared to \lsst alone), 
we expected improved higher information due to the addition of UV photometry beyond \lsst, which is not observed in Figure~\ref{fig:tlmz}. 
We hypothesize that the influence of survey depth on catalog size and redshift distribution may dominate over the effect of the SED features in the UV.

To distinguish between these effects, we generate a new catalog of \castor UV-optical photometry for a fixed catalog size of $\approx 10^5$ galaxies and calculate $\tav(z)$ on the catalog with each pair of filters, leaving the other filter out one at a time, 

i.e.~ when dropping the \castor g-, u-, or uv-band, we use $\{uv, uv-u\}$, $\{uv, uv-g\}$, $\{g, uv-g\}$ respectively to train the approximator to $\tav$. 
The result of this experiment is shown in Figure \ref{fig:drop}, where lower $\tav(z)$ indicates the missing filter was more informative than the remaining ones.

We note that $\tav(z \lesssim 1.5)$ for each pair is comparable and lower than the all-band \castor-only case of Figure~\ref{fig:tlmz}, confirming that the information content is comparable in each pair of filters yet still greater when all three are combined. 

However, $\tav(z\gtrsim 1.5)$ monotonically decreases upon dropping the \castor u or the \castor g bands, while dropping the uv band reduces the potentially recoverable redshift information at $1.5\lesssim z\lesssim 2$ but increases it at $z \gtrsim 2$ compared to dropping the u and the g bands, which makes sense because the constraining power at high redshift is due to the u-uv color.

These results are consistent with the hypothesis that the relative lack of UV information observed in the combined survey cases is due to the brightness of UV sources relative to the depth of the \castor survey and not due to a lack of SED features in the observed frame UV-optical bands.

\begin{figure*}
  \includegraphics[width=0.99\textwidth]{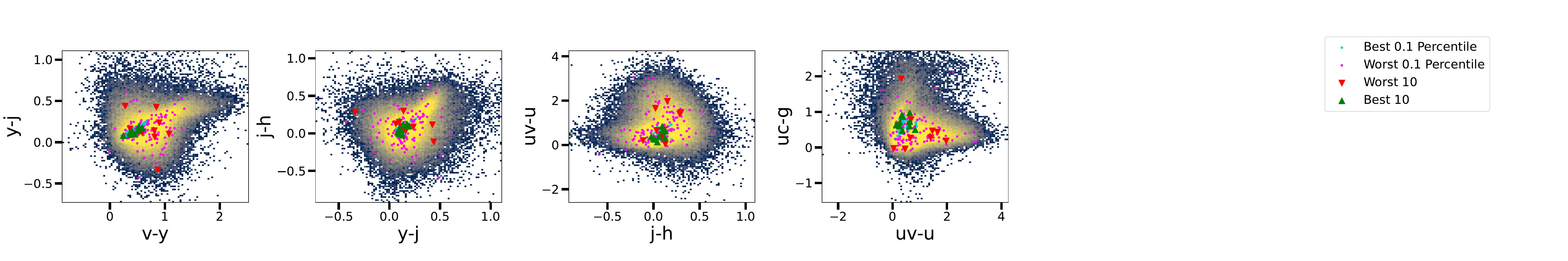}\\
  \includegraphics[width=0.83\textwidth]{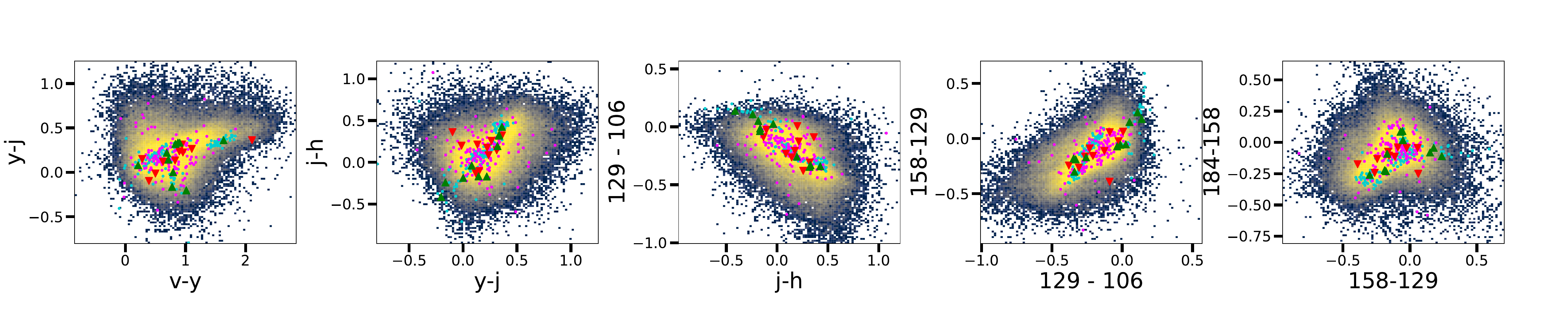}\\
  \includegraphics[width=0.99\textwidth]{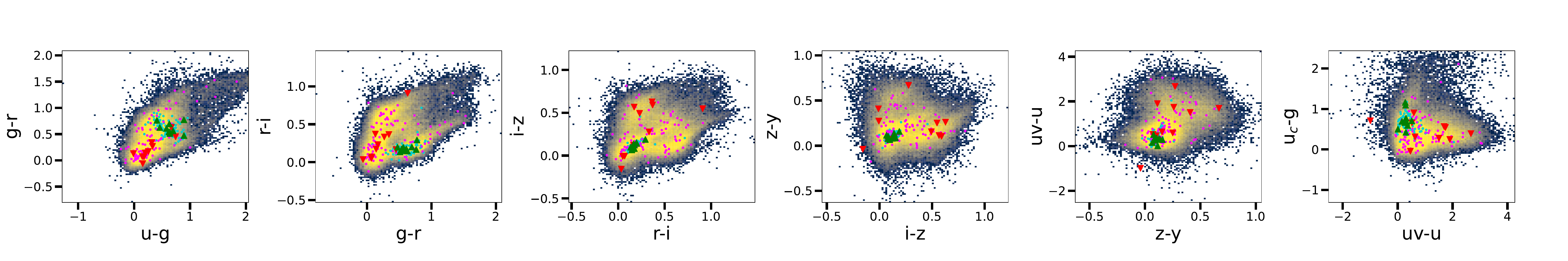}
  \caption{
  Color-color plots for the \euclid+\castor (top), \euclid+\rst (middle), and \lsst+\castor (bottom) survey combinations, highlighting the greatest (cyan) and least (magenta) tenth of a percentile, as well as the top ten (green upward-pointing triangles) and bottom ten (red downward-pointing triangles), in redshift information gain relative to a single survey.
  In most cases, the galaxies that benefit most and least from additional photometry are not visually distinct from the bulk of the population in color-space projections despite being extreme outliers in redshift information content.}
  \label{fig:color-color-delta-tav} 
\end{figure*}

\subsection{Color-dependence of \textit{TheLastMetric}}
\label{sec:tavcolor}

\begin{figure*}
  \centering
  \includegraphics[width=0.9\textwidth]{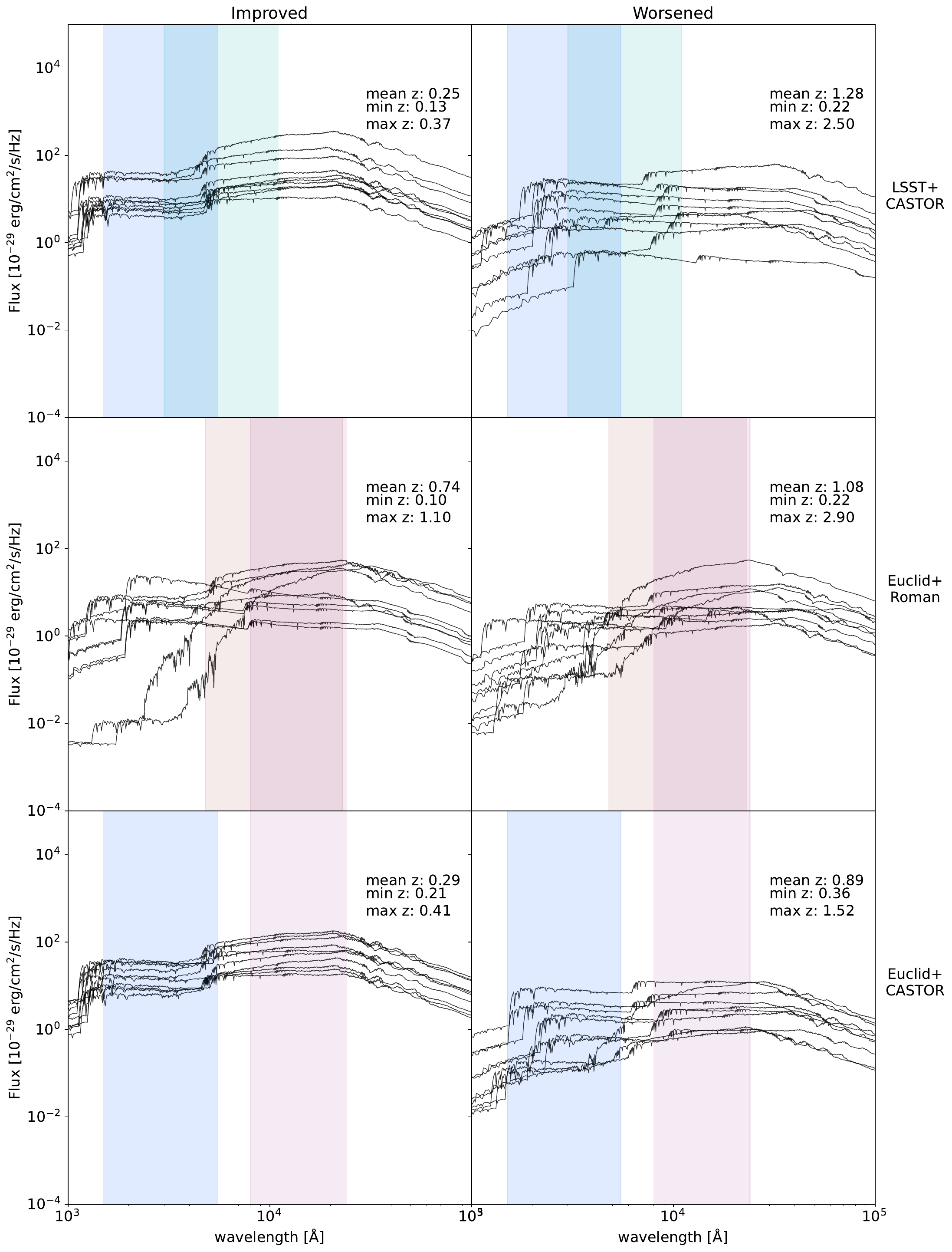}
  \caption{SEDs of galaxies with extreme redshift information gain (left column) or loss (right column) in the \lsst+\castor (top), \euclid+\rst (middle), and \euclid+\castor (bottom) survey combinations, with shaded regions for the bandpasses of each constituent survey. 
  These classes of SEDs share some spectral features despite their disparate locations in the color-color projections of Figure~\ref{fig:color-color-delta-tav}.}
  \label{fig:SEDsbytav} 
\end{figure*}

Figure~\ref{fig:tavdists} raises questions about the individual galaxies that gain or lose redshift information when additional photometric data is made available.
Here, we examine redshift information in color-color space, locating both where in color space is there the most information about redshift and where galaxies with the greatest and least information gain $\Delta I(z, x_{ph})$ (as defined in Equation~\ref{eqn:deltaI}) reside.
This section presents a cursory analysis of two-dimensional projections of the higher-dimensional color space;
a more rigorous quantification of the distance between redshift information-stratified populations in the unprojected color-space is left for future work.

Figure~\ref{fig:color-color-delta-tav} shows color-color plots for the \lsst+\castor, \euclid+\rst, and \euclid+\castor survey combinations, distinguishing the full photometry of the combined survey, the most extreme tenth of a percent thereof as measured by the information gain metric, and the most extreme tenth of a percent as measured just by \textit{TheLastMetric} on the combined photometry. 

While one might expect these extreme populations to be identifiable in features or borders of the color-color distributions, they are instead found in high-density regions of the two-dimensional projections of color-color space.

Intuitively, since photo-$z$s are determined from the differences in adjacent band magnitudes, one would expect that galaxies with zero, or close to zero, color to the most difficult to constrain redshifts, and correspondingly less informative photometry. 
Indeed, we see some evidence of this as the galaxies with the very lowest information gain do have colors scattered around zero in color-color plots with bands split across multiple surveys. 
For example, one can see in the \rst+\euclid (j-h, 129-106) and \lsst+\castor (z-y, uv-u) color-color plots that the bottom tenth of a percentile by information gain tends to be in the vicinity of the origin for one of the two combined surveys and small color values in the other. 

The most striking differences in regions of color space occupied by extreme positive and negative $\Delta I(z, x_{ph})$ are seen in the \lsst+\castor case. 
While galaxies with information loss are scattered throughout the color-color projections, information gain is associated with positive $\{g-r\}$, near-zero $\{g-r, r-i\}$ and positive $\{u_c-g\}$ (the difference in castor u and g band) color, mirroring our previous observation that the \castor u- and g-bands contain important information about redshift that is lost at the population level due to the decline in the number of sources detected at high redshift. 

\euclid+\castor is notable for having the most widely separated filter bandwidths of the combinations considered.
The galaxies with the lowest information gain are non-localized across the (y-j, v-y, j-h) pairwise color-color plots, whereas the galaxies with highest information gain are constrained to a smaller region with small (or near-zero) colors in one or more of the \euclid-only pairs. 
The corresponding color-color plots for the \castor-only bands, however, show that the galaxies with the largest redshift information gain are offset and span a different range in color-space from the \euclid-only color-color plots. 
This provides some intuition for the origin of the wide distribution in the information gain shown in Figure~\ref{fig:tavdists} for this survey combination, as the loss of information is widely distributed in the \castor color-color space while the gain is confined to a smaller region corresponding to small values of uv-u color.  

Since \euclid and \rst are both infrared surveys, we expect the distribution in information gain to be peaked around zero, as much of the information about redshift found in one survey will also be found in the other.

Across several of the color-color plots, we do find that the smallest and largest information gain occupy similar regions of color space and span the full color space of our simulated photometry. 
However, the \euclid-only (v-y, y-j) shows some evidence for a population of enhanced redshift information for small values of (v-y) and large values of (y-j) color, which may correspond to the expected degeneracy breaking in the y-band reported in \cite{Graham_2020}. 
There is some apparent segregation in the \rst-only (158-129, 184-158) color-color plot in the locations of largest and smallest information gain in the combined survey, where information loss is associated with near zero 158-129 color. 
Despite these trends, the most extreme galaxies in terms of redshift information gain or loss are well-distributed across the densest regions of the full color-color spaces.

Overall, this examination demonstrates that the galaxies with largest and smallest information gain are not well-separated in the projected \euclid+\castor and \euclid+\rst color-color spaces, illustrating the utility of the information-theoretic perspective in understanding the relationship between galaxy color and potential quality of redshift estimates. 

Nonetheless, we repeat the caveat that this analysis is limited by the necessity of two-dimensional projections of a higher-dimensional color space.
Future work will quantify the distance and mutual coverage between redshift information-stratified populations in the unprojected color-space.

\subsection{SED stratification by information content}
\label{sec:tavselect}

Based on the point-wise information gain across survey combinations, we can examine SEDs corresponding to parts of color-space that are associated with large changes in the redshift mutual information lower bound. 
As discussed in the preceding sections, increases in the redshift information content of photometry are associated with the breaking of degeneracies in the mapping of SEDs, photometry, and redshift. 
Surprisingly, the point-wise mutual information gain is not positive definite, implying that additional photometry can also, in some cases, add only noise to the distribution of redshift conditioned on photometry; 
\cite{Graham_2020} reported a similar effect where additional \castor photometry was found to increase the precision of a photo-$z$ estimate at the cost of the accuracy of that estimate. 
The following test of our expectation of a correlation between SED features and the redshift information content considers the cases of \lsst+\castor and \euclid+\castor based on the results of Section~\ref{sec:zdep} and \euclid+\rst because \euclid and \rst have similar population-wide redshift information content on a single survey basis but combine to yield a net information gain relative to \euclid-only data.

Figure~\ref{fig:SEDsbytav} shows the SEDs of the ten galaxies with highest and lowest redshift information gain, which were highlighted symbolically in Figure~\ref{fig:color-color-delta-tav}. 
Though the SEDs may have disparate locations in the color-color projections, they do share spectral features that \textit{TheLastMetric} picks up on.
Overall, the galaxies whose redshift information content worsens with the addition of another survey have relatively flat SEDs across multiple survey bands, while the SEDs of the galaxies whose redshift information content improves with more photometry exhibit a greater variety of shapes, featuring strong breaks or wavelength-dependent evolution in single or multiple bands, i.e.~ the signature of degeneracy-breaking.

The most-improved group across \lsst+\castor have mean redshift $\hat{z}\approx 0.2$ and flat SEDs across the \castor bands with the Balmer break in the overlap region with \lsst, corresponding to the expected degeneracy in the Balmer break between \castor and \lsst. 
The story is similar for \euclid+\castor's most-improved galaxies' SEDs, which also have mean redshift $\hat{z}\approx 0.2$ and share features with one another, peaking in the IR bands of \euclid with mild wavelength evolution across the IR and a strong break at the edge of \castor's UV coverage, again indicating the improvement is due to localization of the Balmer break in the \castor u and g. 
Unsurprisingly given the degree of overlap in survey bandpasses between \euclid and \rst, the SEDs of the most-improved galaxies in this survey combination have more diversity compared to the other survey combinations, with a large degree of broadband evolution across the survey bandpasses, with some galaxies strongly peaking in the most extreme \rst band and others having flat or inverse evolution with increasing wavelength, and higher redshifts overall (aside from one outlier with $z \approx 0.1$).

Galaxies that appear to lose redshift information when more photometric bands are included can be understood as having their redshift information diluted by the addition of uninformative photometry. 
For \lsst+\castor, this group is characterized by more diversity than its most-improved group, with a wide range in redshift from $0.2 \lesssim z \lesssim 2.5$ and SEDs with their Balmer break in the \lsst bandpasses and in some cases a Lyman break visible at the blue end of the \castor bandpasses, suggesting a physical degeneracy that can't be broken without further wavelength coverage.
Aside from those features, the \lsst+\castor galaxies exhibit a mix of broadband evolution peaking in the near-infrared and flat evolution across the \lsst and \castor bands, meaning we would not expect \castor's deeper UV coverage to break a degeneracy

Nonetheless, several of the \lsst+\castor SEDs with extreme information loss have both the Lyman and Balmer breaks detected in the combined survey, with negative (decreasing flux with increasing wavelength) evolution of the SED seemingly associated with information loss across the combined photometry.  

As anticipated, we see a similar diversity of typology and redshift among the SEDs with extreme redshift information loss under the \euclid+\rst survey combination, with a large degree of broadband evolution featuring positive slopes in the \euclid bandpasses (peaking in the IR) and large breaks at wavelengths blue-ward of the available filters, with some having flatter IR evolution across \rst or shallower breaks if they're covered by the filters at all.   

In the same vein, the \euclid+\castor galaxies that see redshift information loss with additional photometry are associated with either flat evolution across or a Lyman break detected at the edge of the \castor filters and a flat SED in \euclid, leaving the Balmer break in the unobserved region between their wavelength coverage. 
The flatness of SEDs associated with information loss in a combined survey is consistent with the intuition that uninformative photometry corresponds to colors that don't capture SED evolution with wavelength, where a non-detection of the Balmer/Lyman Break is a particular example of non-detected SED evolution.

\section{Conclusions} \label{sec:conclusions}

In this work, we have forecasted the quality of photometric redshift recovery of the Stage IV galaxy surveys, \euclid, \rst, and \lsst, and the \castor UV survey using \textit{TheLastMetric}, $\tav$, which quantifies the potentially recoverable redshift information content of photometry.

We show that \textit{TheLastMetric} is richly informative without favoring any particular science use case of photo-$z$ data products and without committing to the choice of any specific photo-$z$ estimator nor its requisite priors.
Though our demonstration is performed in the context of a simulated catalog that itself constitutes a prior, as does the implementation of the approximator of the probability space of redshift and photometry, this work nonetheless exemplifies a consistent and general approach to the problem of combining photometry across surveys with very different selection functions and strategies.

When each Stage IV survey is considered in isolation, \lsst is most informative across all redshifts $0\leq z\leq 3$.
While little information is to be gained by the addition of other survey data at low redshift,

the inclusion of IR photometry from \euclid and \rst significantly improves the redshift information content due to detection of the Balmer break. 
Although earlier work indicated UV photometry would be uninformative of redshift at $z\ge 1.7$, we find that the impact of breaking the Lyman-Balmer degeneracy is suppressed in population-level metrics due to the smaller number of sources that satisfy combined UV, optical, and infrared magnitude limits rather than the presence of absence of observer-frame UV spectral features. 

Given the important role of survey depth, we thus expect \castor to reach an upper limit on the ability of upcoming UV surveys, such as ATLAST and UV-Explorer, to contribute to combined survey photometric redshift recovery.

In the latter half of this work, we used the redshift evolution in order to study how infrared, optical, and UV photometry provide differing information based on the position of SED features. 
A difference in mutual information effectively defines a notion of information gain of a survey combination relative to its constituents, which we study as a function of high-dimensional color before  examining SEDs stratified by information gain.
 
We observe some anticipated trends in the SEDs whose potentially recoverable redshift information is most sensitive to additional photometry but 

leave a more thorough exploration of the redshift information content of galaxy SEDs and populations to future work.

The scope of this paper is, admittedly, small, but \textit{TheLastMetric} and its extensions introduced herein may be influential in addressing other aspects of survey design.
The redshift information content of a survey (or combination thereof) can instead be used to optimize decisions about the choice of wavelength and filter coverage, depth of overlapping survey footprints, and cadence (where differential filter depth can have poorly characterized effects on early science). 
Estimator-independent parameter recovery assessment could also be of value for other astrophysical or cosmological properties beyond redshift, such as galaxy mass or star formation rate. 

Future work may develop an information-theoretic metric for components of cosmological probes, such as the redshift distribution $n(z)$ derived from galaxy clustering statistics, or for end-to-end cosmological parameter constraints from photometry and imaging.

\section*{Acknowledgements}

We thank the \castor team for welcoming, inspiring, and encouraging our work as part of their mission planning activities. 
BRS is supported by the \lsst Discovery Alliance Data Science Fellowship Program, which is funded by \lsst-DA, the Brinson Foundation, and the Moore Foundation.
AIM is supported by Schmidt Sciences. 
RS is supported by funding from the Canadian Space Agency and the National Research Council of Canada.
We thank John Franklin Crenshaw for helpful conversations.

\bibliography{sample631}{}
\bibliographystyle{aasjournal}

\end{document}